\documentclass{article}

\usepackage{arxiv}

\usepackage[utf8]{inputenc}
\usepackage[T1]{fontenc}
\usepackage{amsmath}
\usepackage{amssymb}
\usepackage{graphicx}
\usepackage{booktabs}
\usepackage{tabularx}
\usepackage{subcaption}
\usepackage[numbers,sort&compress]{natbib}
\usepackage{microtype}
\usepackage{xcolor}
\usepackage[hidelinks]{hyperref}

\title{Regime-Aware Coupled Orchestration of Neural Operator Specialists for Multiphysics Dynamics}

\author{
Paolo Guida\thanks{Corresponding author.} \\
Clean Energy Research Platform \\
King Abdullah University of Science and Technology \\
Thuwal 23955, Saudi Arabia \\
\texttt{paolo.guida@kaust.edu.sa} \\
\And
Vijayamanikandan Vijayarangan \\
Clean Energy Research Platform \\
King Abdullah University of Science and Technology \\
Thuwal 23955, Saudi Arabia \\
\texttt{vijayamanikandan.vijayarangan@kaust.edu.sa} \\
\And
Hong G. Im \\
Clean Energy Research Platform \\
King Abdullah University of Science and Technology \\
Thuwal 23955, Saudi Arabia \\
\texttt{hong.im@kaust.edu.sa} \\
\And
William L. Roberts \\
Clean Energy Research Platform \\
King Abdullah University of Science and Technology \\
Thuwal 23955, Saudi Arabia \\
\texttt{william.roberts@kaust.edu.sa} \\
}

\begin{document}
\maketitle

\begin{abstract}
Real-world physics often unfolds over a variety of characteristic lengths and time scales. These non-uniformities are among the main reasons numerical simulations remain expensive when used predictively on practical systems: the same computational domain can contain smooth transport, vortical structures, thin interfaces, reaction fronts, and rapidly relaxing regions. Neural operators are attractive surrogates for such problems. However, a single architecture is still expected to represent all these local behaviours with a single set of weights and a single inductive bias. In this work, we introduce the regime-aware coupled neural orchestrator (RCNO), a framework that learns to distribute and connect a library of local operator specialists in space and time. The bank contains Fourier Neural Operator (FNO), Laplace Neural Operator (LNO), and DeepONet models trained at low, intermediate, and high levels of local activity. For every patch, an orchestrator builds a causal query from the current fields, neighbouring halo values, local regime features, patch identity, activity level, and case parameters. The nine expert embeddings exchange two rounds of query-conditioned message passing before an edge-biased cross-attention stage produces a dispatch distribution that can be evaluated sparsely. Local predictions are assembled into a global field and projected onto system-specific physical constraints. A seam-aware extension additionally learns residual corrections across internal patch interfaces. We apply the same orchestration architecture to two qualitatively different systems: reacting flow and Cahn--Hilliard--Navier--Stokes (CHNS) dynamics. The results show that useful operator specialisation exists across local regimes, that the advantage of conditional composition increases as the spatial decomposition becomes finer, and, on reacting flow, that explicitly treating patch interfaces improves continuity, PDE consistency, and rollout accuracy. These results support neural-operator orchestration as a modular alternative to applying one surrogate uniformly across heterogeneous multiphysics domains.
\end{abstract}
 
 
\section{Introduction}
\label{sec:intro}
 
Mathematical representations of complex flow phenomena are central to understanding, forecasting, and controlling a wide range of natural and engineering processes in biology, atmospheric chemistry, combustion, and multicomponent and multiphase transport. These complex nonlinear dynamical systems are described by sets of partial differential equations (PDEs), and simulating them, particularly at scale, is extremely challenging because the constituent physical processes coexist and interact over a wide range of characteristic scales. In reactive flows, for instance, chemical reactions, whose rates often depend exponentially on temperature, influence temperature and composition fields that follow diffusive dynamics around thin reaction fronts. Multiphase flows present similar challenges: the interface between phases concentrates gradients over a narrow region while the bulk phases remain comparatively smooth. In practical systems these mechanisms operate together. A spray combustor, for instance, simultaneously combines turbulence, gas-liquid interfaces, transport, and chemical reactivity \cite{meierhofer2021synthesis,lefebvre2017atomization}, and the same applies to electrolysers, where the applied potential drives the evolution of hydrogen bubbles \cite{li2018situ,falcao2020review}.
 
Researchers have been developing a variety of strategies to reduce the cost of high-dimensional nonlinear simulation, including empirical multiscale modelling, reduced representations, data-driven surrogates, and data-driven numerical models \cite{rabczuk2023machine,zhu2022review,gao2024multiscale}. Neural operators (NOs) are particularly attractive in this setting. In fact, NOs, once trained, can be reused across families of parameterised PDE solutions \cite{kovachki2023neural,azizzadenesheli2024neural} since they are built to learn how to map from function spaces. 
 
Among NOs, Fourier neural operators (FNOs) learn global spectral mappings \cite{li2021fno,duruisseaux2025fourier}, and Laplace neural operators (LNOs) use a pole-residue representation in the Laplace domain \cite{cao2024laplace}. Another popular NO architecture is DeepONet, which separates the input and output into a complementary branch-trunk representation \cite{lu2021deeponet}. Each of these architectures may perform better in some regimes and worse in others. Moreover, although neural operators can represent multiple regimes, their accuracy can deteriorate when the learned dynamics must span strongly heterogeneous or extrapolative conditions \cite{weng2025extended,nath2026amore}. The opportunity to use dedicated surrogates for specific regimes or variables has been exploited in multi-operator, codomain-attention, and mixture-of-experts approaches that introduce modularity to reduce interference across heterogeneous PDE tasks or physical regimes \cite{rahman2024pretraining,deighan2025mixture,wang2025moepot,feng2026dynamic,sharma2026eradicating,wang2026latent}. Related operator-learning work has also highlighted the particular difficulty of sharp, discontinuous, or strongly localised solution features \cite{cavallazzi2026walsh,lin2023operator,dang2026smooth,guida2026data}.
 
We propose a learned orchestrator for specialised neural operators and attempt to demonstrate that a collection of independently trained neural operators can be dynamically assembled over a physical domain so that the global surrogate follows the governing multiphysics dynamics with higher accuracy than fixed local-operator policies. We name our framework regime-aware coupled neural orchestrator (RCNO). RCNO separates the learning problem into two stages. First, local neural operator specialists are trained independently on different local dynamical regimes and then frozen. Second, a graph orchestrator learns which specialist should act on each spatial patch at each time step and how the expert representations should interact before dispatch. The selected local predictions are assembled, physically projected, and fed back autoregressively.
 
The orchestration takes place via query-conditioned message passing, so that each specialist's representation depends on the local state under consideration. The target of the training process is not only the minimisation of the trajectory error but also the reduction of the PDE defect, which enters through the physics-aware expert ranking and makes the procedure \emph{de facto} physics-informed.  We also introduce a "stitching" process that implements an operator that learns to reconcile patch boundaries, minimising spurious discontinuities. We evaluate the framework on a reacting flow and on a two-phase system described by the Cahn--Hilliard--Navier--Stokes (CHNS) equations. Both are instantiated within the same orchestration framework, with each system supplying its own state variables, physical parameters, local activity descriptors, PDE residuals, and projections. 

The rest of the manuscript is organized as follows: Section~\ref{sec:method} describes the problem formulation and methodology employed in this study. The results were organised into two sections: Section ~\ref{sec:rf} discusses the results from the reactive flow configuration, and Section ~\ref{sec:chns} discusses the multiphase flow configuration. Finally, section ~\ref{sec:conclusion} summarizes the work.
%The work proceeds with the methodology (Section~\ref{sec:method}), the two applications (Sections~\ref{sec:rf} and~\ref{sec:chns}), and conclusions (Section~\ref{sec:conclusion}).
 
\section{Methodology}
\label{sec:method}
 
\paragraph{Problem definition}
 
In this work we propose a method to approximate the dynamic evolution of a state $\mathbf q(\mathbf x,t)$:
\begin{equation}
  \partial_t \mathbf q = \mathcal F(\mathbf q;\boldsymbol\theta),
  \qquad \mathbf x\in\Omega,
  \label{eq:pde-general}
\end{equation}
where $\mathbf q(\mathbf x,t)\in\mathbb R^{d_q}$ denotes the state vector at spatial location $\mathbf x$ and time $t$, $\partial_t\mathbf q$ is its local time derivative, and $\mathcal F$ is the nonlinear spatial differential operator containing the governing transport, source, reaction, and/or capillary terms. The case-parameter vector $\boldsymbol\theta\in\mathbb R^{d_\theta}$ stores the physical parameters that vary across trajectories, such as transport coefficients, dimensionless groups, interface parameters, or reaction-rate coefficients. Finally, $\mathbf x\in\Omega\subset\mathbb R^d$ denotes the spatial coordinate, and throughout this work we use $d=2$.
 
Neural operators approximate mappings between function spaces \cite{lu2021deeponet,kovachki2023neural} and have been used broadly as surrogate solvers for parameterised dynamical systems \cite{li2021fno,azizzadenesheli2024neural}. The neural operator $\mathcal G$ is trained to advance the state $\mathbf q$ by a time step $\Delta t$,
\begin{equation}
  \mathcal G:\;(\mathbf q_t,\boldsymbol\theta)\longmapsto \mathbf q_{t+\Delta t},
  \label{eq:global-no}
\end{equation}
and is applied autoregressively, $\mathbf q_{t+n\Delta t} = \mathcal G\big(\mathbf q_{t+(n-1)\Delta t},\boldsymbol\theta\big)$, to predict an arbitrary number of steps ahead. The complexity of this task grows sharply when the local flow map combines smooth transport, relaxation, vortical motion, and sharp or stiff dynamics \cite{cavallazzi2026walsh,dang2026smooth,weng2025extended,nath2026amore}.
 
\paragraph{Regime-aware coupled neural orchestrator}
To prove the concept of routing neural operator specialists in real time, we limit our analysis to a physical domain $\Omega$ divided into $P$ non-overlapping patches. We propose to substitute the operator $\mathcal G$ with a composition of routed local specialists,
\begin{equation}
  \widehat{\mathbf q}_{t+\Delta t}
  =
  \mathcal C\!\left[
    \mathcal S\!\left(
      \sum_{p=1}^{P} \mathcal P_p^\top
      \left(
        \sum_{k=1}^{K}\pi_{p,k}^{t}\,
        \mathcal N_k\big(
          \mathcal P_p\widehat{\mathbf q}_{t},
          \mathcal H_p\widehat{\mathbf q}_{t},
          \boldsymbol\theta
        \big)
      \right)
    \right)
  \right],
  \label{eq:coupled-update}
\end{equation}
where $\mathcal P_p$ extracts patch $\Omega_p$ from the domain $\Omega$ and $\mathcal P_p^\top$ places it back, $\widehat{\mathbf q}_{t}$ is the predicted global state with $\widehat{\mathbf q}_0 = \mathbf{q}_0$, and $\mathcal H_p$ extracts the halo of patch $p$, i.e.\ the single row or column of cells of the four neighbouring patches (north, south, east, and west) that is adjacent to $\Omega_p$. The symbol $\mathcal N_k$ denotes the $k$-th local specialist operator, and $\pi_{p,k}^{t}\ge 0$ is the dispatch weight assigned to specialist $k$ on patch $p$ at time $t$, with $\sum_k\pi_{p,k}^{t}=1$. We analyse two modes. In the "single best expert" mode, $\pi_{p,k}^{t}=1$ for the highest-scoring specialist and $0$ otherwise, so that only one operator is evaluated per patch; in the averaged mode, all candidates are blended according to their dispatch weights. Unless stated otherwise, all reported results use the "single best expert" mode.
After composition, the learned seam-correction operator $\mathcal{S}[\cdot]$ reduces inconsistencies between neighbouring patches, and the physical projection operator $\mathcal{C}[\cdot]$ imposes system-specific hard constraints. In the reacting case, for instance, $\mathcal C$ enforces non-negative species mass fractions that sum to one, together with temperature and vorticity bounds.
 
The term \emph{coupled} refers to the two mechanisms that make the local predictions interdependent: the expert representations exchange messages conditioned on the local query before dispatch, and neighbouring patches are coupled through the halo inputs and the seam correction $\mathcal S$.
 
The algorithm therefore proceeds by dividing the computational domain into patches and their halos. For every patch, RCNO builds a router query vector $\mathbf z_p^t$ (Eq.~\eqref{eq:router-input}) that is mapped by a learned linear layer to a contextual token $\mathbf h_p^t\in\mathbb R^{d_h}$. The token conditions message passing on a graph whose nodes are the $K$ experts of the bank $\mathcal E$ (Eq.~\eqref{eq:expert-bank}), and an attention stage then assigns a scalar cost $\gamma_{p,k}^t$ to each candidate expert. The costs are converted into dispatch weights through a softmax,
\begin{equation}
  \pi_{p,k}^{t}=\frac{\exp\!\left(-\gamma_{p,k}^{t}\right)}{\sum_{k'=1}^{K}\exp\!\left(-\gamma_{p,k'}^{t}\right)},
  \label{eq:dispatch-softmax}
\end{equation}
so that a lower cost corresponds to a higher probability that expert $k$ fits patch $p$. Once the experts have been evaluated and assembled, the remaining steps are seam correction through $\mathcal S$ and physical projection through $\mathcal C$. The resulting field $\widehat{\mathbf q}_{t+\Delta t}$ is fed back autoregressively as the input for the subsequent time step.
 
\paragraph{Specialists}
 
In our framework, local specialists are trained to predict the full physical state vector simultaneously. In reacting flow, for instance, each expert receives and processes temperature, vorticity, and species fields together on its local patch. As previously stated, we limit our analysis to two-dimensional cases on a regular grid, whose domain $\Omega$ is divided into $P_x\times P_y$ (in total $P$) non-overlapping rectangular patches,
\begin{equation}
    \Omega=\bigcup_{p=1}^{P}\Omega_p,\qquad \Omega_p\cap\Omega_{p'}=\varnothing\quad (\forall p\neq p').
    \label{eq:patch-decomp}
\end{equation}
Every expert is position-agnostic, allowing it to be evaluated on any patch. Local boundary ambiguity is reduced by supplying each expert with the normalised current patch
\begin{equation}
  \widetilde{\mathbf q}_p^t=\big(\mathcal P_p\widehat{\mathbf q}_t-\bar{\mathbf q}\big)\oslash\boldsymbol\sigma_q,
  \label{eq:patch-normalisation}
\end{equation}
where $\bar{\mathbf q}$ and $\boldsymbol\sigma_q$ are per-variable training-set means and standard deviations and $\oslash$ denotes element-wise division, together with the halo values $\mathcal H_p\widehat{\mathbf q}_t$ normalised in the same way, local coordinates centred in $[-1,1]^2$, and normalised case parameters $\widetilde{\boldsymbol\theta}$. All expert architectures share the same residual convention,
\begin{equation}
  %\widetilde{\mathbf{q}}_{p,\tau+1} = \widetilde{\mathbf{q}}_{p,\tau} + \sum_{k=1}^K \alpha_{p,k,\tau} \, \Delta\widetilde{\mathbf{q}}_{p,k,\tau}, \qquad 
  \widehat{\mathbf{q}}_{p,\tau+1} = \max\Big(\mathbf{0}, \, \bar{\mathbf{q}} + \boldsymbol{\sigma}_q \odot \widetilde{\mathbf{q}}_{p,\tau+1}\Big),
  \label{eq:residual_expert}
\end{equation}
which gives every architecture the same identity shortcut when the local one-step map is close to the identity and makes heterogeneous expert outputs directly assemblable, because each is returned through the same normalisation.
 
In this application, we limit the expert pool to three operator architectures and three local activity levels,
\begin{equation}
  \mathcal E =
  \{\mathrm{FNO},\mathrm{LNO},\mathrm{DeepONet}\}
  \times
  \{\mathrm{smooth},\mathrm{mixed},\mathrm{sharp}\},
  \qquad K=|\mathcal E|=9.
  \label{eq:expert-bank}
\end{equation}
Each of the nine specialists is pretrained on patches from its own activity level and then frozen; the subsequent composition stage trains only the orchestrator and the seam-correction operator $\mathcal S$. The architectural labels denote different inductive biases rather than fixed physical roles: FNO performs learned low-mode Fourier mixing \cite{li2021fno,duruisseaux2025fourier}, LNO uses a Laplace-domain pole--residue representation \cite{cao2024laplace}, and DeepONet combines a branch representation of the input function with a coordinate-dependent trunk \cite{lu2021deeponet}. The activity label records the regime a specialist was trained on, and the routing stage remains free to select any architecture at any activity level.
 
At every saved time, each patch is assigned a feature vector $\mathbf r_p^t$, computed from the current state alone so that it remains available throughout autoregressive prediction. A scalar activity $a_p^t$ is constructed from the selected components of $\mathbf r_p^t$ and divided into three levels using two thresholds $\tau_1$ and $\tau_2$, the pooled training-set activity values, estimated once and held fixed thereafter,
\begin{equation}
  \ell_p^t =
  \begin{cases}
    0, & a_p^t<\tau_{1},\\
    1, & \tau_{1}\le a_p^t<\tau_{2},\\
    2, & a_p^t\ge\tau_{2}.
  \end{cases}
  \label{eq:levels}
\end{equation}
The terms smooth, mixed, and sharp, therefore, denote low, intermediate, and high measured activity, and they should be read as empirical dynamical buckets rather than ground-truth phases. Both $\mathbf r_p^t$ and $\ell_p^t$ are recomputed from the predicted state during rollout, which allows the local regime of a patch to change as a front, an interface, or a vortical structure moves through the domain.
 
\paragraph{Graph orchestrator}
\label{sec:router}
 
The router query of patch $p$ at time $t$ concatenates
\begin{equation}
  \mathbf z_p^t=\Big[\,
  \mathbf s\big(\widetilde{\mathbf q}_p^t\big),\;
  \bar{\mathbf h}_p^t,\;
  \mathbf r_p^t,\;
  \mathbf e_p,\;
  \mathbf e_{\ell_p^t},\;
  \widetilde{\boldsymbol\theta}
  \,\Big],
  \qquad
  \mathbf h_p^t=\mathbf W_z\mathbf z_p^t+\mathbf b_z\in\mathbb R^{d_h},
  \label{eq:router-input}
\end{equation}
where $\mathbf s(\cdot)$ collects per-channel patch statistics, $\bar{\mathbf h}_p^t$ contains the mean normalised halo values on each of the four sides, $\mathbf e_p\in\{0,1\}^P$ is the one-hot patch identity, $\mathbf e_{\ell_p^t}\in\{0,1\}^3$ is the one-hot activity level, and $d_h$ is the embedding width.
 
Each expert $k$ is represented by a learned embedding $\mathbf u_k^{(0)}\in\mathbb R^{d_h}$. The experts form the nodes of a relation graph whose edges connect experts who share an architecture or an activity level, with $\mathcal N(k)$ denoting the neighbours of $k$. For $r=0,\ldots,R-1$ (with $R=2$), the embeddings are updated by query-conditioned message passing,
\begin{equation}
  \mathbf u_{k}^{(r+1)}
  =
  \mathbf u_{k}^{(r)}
  +
  \phi_r\!\Big(
    \mathbf u_{k}^{(r)},\;
    \tfrac{1}{|\mathcal N(k)|}\textstyle\sum_{k'\in\mathcal N(k)}\psi_r\big(\mathbf u_{k'}^{(r)},\mathbf e_{kk'}\big),\;
    \mathbf h_p^t
  \Big),
  \label{eq:message-passing}
\end{equation}
where $\phi_r$ and $\psi_r$ are small multilayer perceptrons and $\mathbf e_{kk'}$ is a learned edge-type embedding. Setting $R=0$ recovers static expert embeddings. An edge-biased cross-attention score between the query and each expert then obtains the candidate cost,
\begin{equation}
  \gamma_{p,k}^t
  =
  -\frac{\big(\mathbf W_Q\mathbf h_p^t\big)^{\!\top}\big(\mathbf W_K\mathbf u_k^{(R)}\big)}{\sqrt{d_h}}
  +\beta_{\ell_p^t,k},
  \label{eq:attention-cost}
\end{equation}
which has the form of scaled dot-product attention without a value projection, since only the scores are needed for dispatch, and $\beta_{\ell,k}$ is a learned bias between activity level $\ell$ and expert $k$.
 
\paragraph{Loss function}
The graph orchestrator and seam-correction operator are trained end-to-end by minimising a composite objective combining physics-aware expert ranking, multi-step autoregressive rollout error, and interface continuity:
\begin{equation}
  \mathcal{L}_{\mathrm{total}} = \mathcal{L}_{\mathrm{rank}} + \mathcal{L}_{\mathrm{roll}} + \lambda_{\mathrm{interface}} \mathcal{L}_{\mathrm{interface}}.
  \label{eq:total-loss}
\end{equation}
The ranking term uses an exhaustive oracle: on training data, every expert is evaluated on every patch, and its one-step cost is
\begin{equation}
  c_{p,k}^t
  =
  \left\|\frac{\widehat{\mathbf q}_{p,k}^{\,t+\Delta t}-\mathcal P_p\mathbf q_{t+\Delta t}}{\boldsymbol\sigma_q}\right\|_{\mathrm{RMS}}^2
  +\lambda_{\mathcal D}\,\mathcal D_{p,k}^{t}
  +\lambda_{\mathrm{clip}}\,\kappa_{p,k}^{t},
  \label{eq:oracle-cost}
\end{equation}
where $\mathcal D_{p,k}^t$ is the patch-local PDE defect for the corresponding benchmark and $\kappa_{p,k}^t$ is the fraction of cells modified by the projection $\mathcal C$. The router costs are regressed onto the oracle costs with a Huber loss,
\begin{equation}
  \mathcal L_{\mathrm{rank}}
  =
  \operatorname*{mean}_{p,k,t}\;
  \rho_\delta\!\left(\gamma_{p,k}^t-\widetilde c_{p,k}^{\,t}\right),
  \qquad
  \rho_\delta(e)=
  \begin{cases}
    \tfrac12 e^2, & |e|\le\delta,\\[2pt]
    \delta\big(|e|-\tfrac12\delta\big), & |e|>\delta,
  \end{cases}
  \label{eq:rank-loss}
\end{equation}
where $\widetilde c_{p,k}^{\,t}$ is the oracle cost standardised over the $K$ experts of the patch and $\delta$ is the Huber threshold. To mitigate compounding error during extended time integration, the framework is also trained through differentiable $H$-step rollouts evaluated on normalised state variables,
\begin{equation}
  \mathcal{L}_{\mathrm{roll}} = \frac{1}{H} \sum_{h=1}^{H} \left\| \frac{\widehat{\mathbf{q}}_{t+h\Delta t} - \mathbf{q}_{t+h\Delta t}}{\boldsymbol{\sigma}_q} \right\|_2^2.
  \label{eq:roll-loss}
\end{equation}
Finally, trace discontinuities across shared internal patch boundaries are penalised through the following:
\begin{equation}
  \mathcal{L}_{\mathrm{interface}} = \operatorname{mean}_{(p,p')}\, J_{pp'}^2 
  \label{eq:interface-loss}
\end{equation}
 where the patches $p$ and $p'$ are immediate neighbours. 
\section{Experiment I: reacting flow}
\label{sec:rf}
 
\paragraph{Benchmark and local specialisation}
 
The first benchmark for testing the RCNO framework is a reactive flow. The state is represented by:
\begin{equation}
  \mathbf q_{\mathrm{RF}}=(\omega,T,Y_1,\ldots,Y_{N_s}).
  \label{eq:rf-state}
\end{equation}
Here, $\omega$ represents vorticity, $T$ temperature, and $Y_i$ the mass fraction of the $i$-th species. The governing equations are

\begin{equation}
    \begin{aligned}
        \partial_t \omega &= -\mathbf u\cdot\nabla\omega + \kappa_{\omega} \nabla^2 \omega + \phi \partial_x T, \\
        \partial_t T &= -\mathbf{u}\cdot\nabla T + \kappa_T \nabla^2T + \phi_T, \\
        \partial_t Y_i &=  -\mathbf{u}\cdot\nabla Y_i + \kappa_Y \nabla^2 Y_i + \phi_{Y_i}, \quad \forall 1 \leq i \leq N_s.     
    \end{aligned}
\end{equation}

Here $\phi_T$ is the heat release term, $\phi_{Y_i}$ is the $i^{th}$ species reaction rate, $\phi$ is the thermal expansion coefficient,  and $\kappa_{-}$'s are the diffusivity coefficients of the corresponding equations.
This case is of particular interest as temperature and species follow advective--diffusive equations with reduced stiff kinetics, a setting in which learned surrogates are known to be particularly sensitive to stiffness and struggle to extrapolate \cite{goswami2024learning,weng2025extended,nath2026amore}. 

\begin{figure}[htbp]
    \centering
    % First Subfigure
    \begin{subfigure}[b]{0.45\textwidth}
        \centering
        \includegraphics[width=\textwidth]{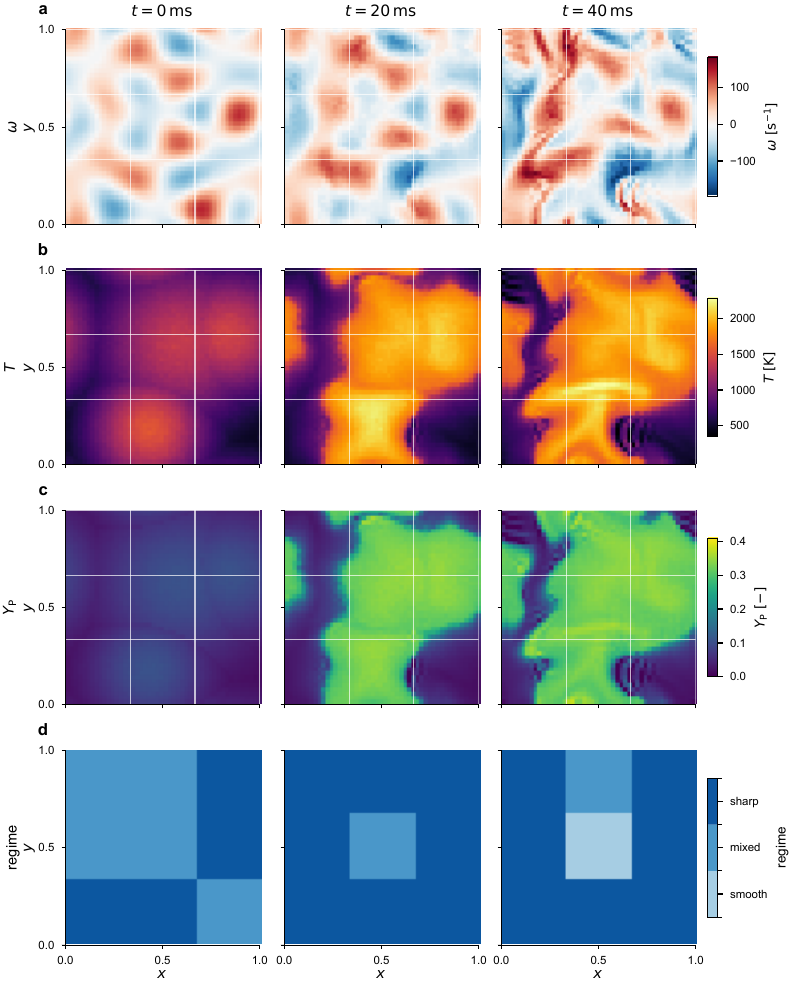}
    \end{subfigure}
    \hfill % Adds horizontal space to push the subfigures to the edges
    % Second Subfigure
    \begin{subfigure}[b]{0.45\textwidth}
        \centering
        \includegraphics[width=\textwidth]{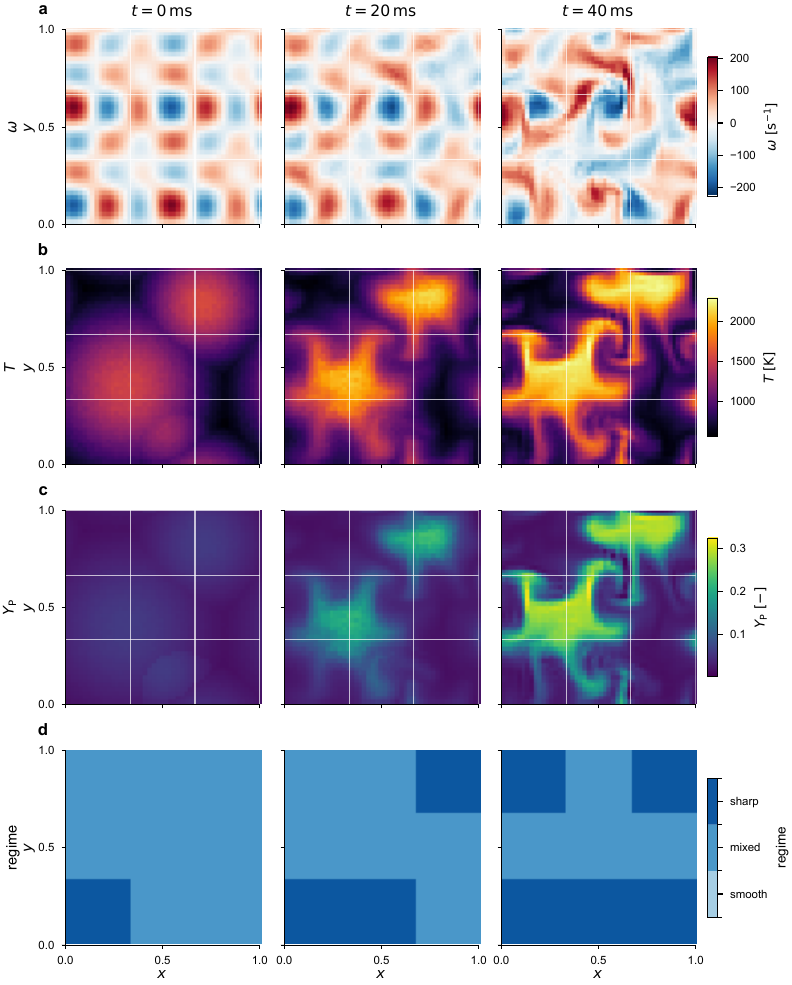}

    \end{subfigure}
    
    \caption{\textbf{Reactive-flow benchmark.} Fast (left) and slow (right) dynamics. Rows $a$--$c$ show vorticity, temperature, and product-species mass fraction, respectively; row $d$ shows the regime assigned to each patch.}
    \label{fig_1}
\end{figure}

The dataset was created by sampling cases with varying chemistry and initial front sharpness to span chemical responses with different characteristic times and both sharp and diffuse spatial distributions. We solved the system using a standard 2D reactive CFD solver suitable for combustion problems, in which chemical reactivity and the flow field are coupled via operator splitting \cite{cuoci2013numerical}. Figure~\ref{fig_1} shows representative fast- and slow-reactivity cases, including vorticity, temperature, product mass fraction, and the regime classification adopted in this work. The next step was to train the ``experts'', neural operators that learn the dynamics over individual patches. The three neural operator architectures selected for the study were FNO, DeepONet, and LNO. All three were evaluated across all patches and regimes and were trained for autoregressive prediction of the system dynamics. The number of cases and additional dataset information about our reaction in a doubly-periodic square domain are reported in Table~\ref{tab_1} and their simplified chemical mechanism in Table ~\ref{tab_reaction_mechanism}.

\begin{table}[t]
\centering
\caption{\textbf{Reactive-flow dataset.} State variables, dataset split, and spatial resolutions used in the reacting-flow benchmark.}
\label{tab_1}
\small
\begin{tabular}{ll}
\toprule
Property & Value \\
\midrule
State variables &
$\omega,T,Y_F,Y_O,Y_I,Y_P,Y_N$ \\
Number of fields & 7 \\
Number of cases & 60 \\
Train/validation/test & 42/9/9 cases \\
Grid resolutions & $48\times48$ and $96\times96$ \\
\end{tabular}
\end{table}

\begin{table}[t]
    \centering
    \caption{Two-step consecutive chemical mechanism.}
    \label{tab_reaction_mechanism}
    \begin{tabular}{c c c}
        \toprule
        Reaction & Reaction rate & Arrhenius term \\
        \midrule
        $F + O \rightarrow I$
        & $\phi_1 = k_1[F][O]$
        & $k_1 = A_1\exp\left(-\frac{E_{a,1}}{RT}\right)$ \\[6pt]
        
        $I \rightarrow P$
        & $\phi_2 = k_2[I]$
        & $k_2 = A_2\exp\left(-\frac{E_{a,2}}{RT}\right)$ \\
        %\bottomrule
    \end{tabular}
\end{table}

\begin{figure}[h!]
  \centering

    \includegraphics[width=0.5\linewidth]{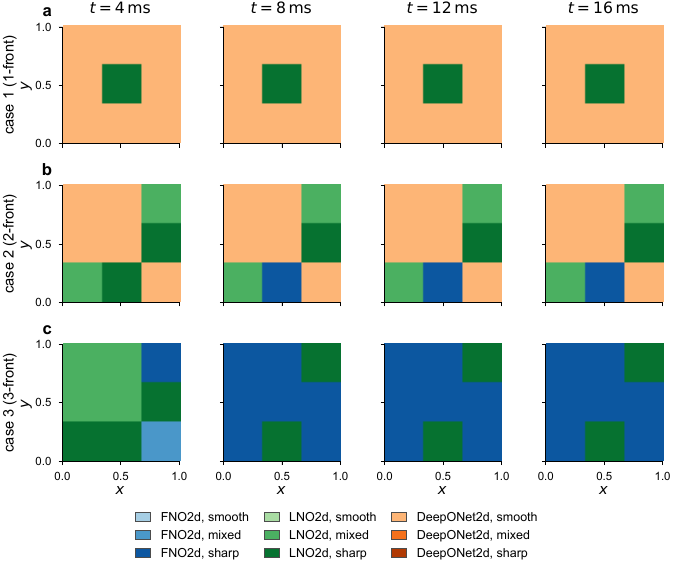}
    \caption{\textbf{Reactive-flow expert dispatch.} Specialist assignments for three test cases on a $48\times48$ grid with $3\times3$ patches of $16\times16$ cells at $t=4,8,12,16$~ms.}
    \label{fig_2}
\end{figure}

\begin{figure}[h!]
  \centering
  \includegraphics[width=0.5\linewidth]{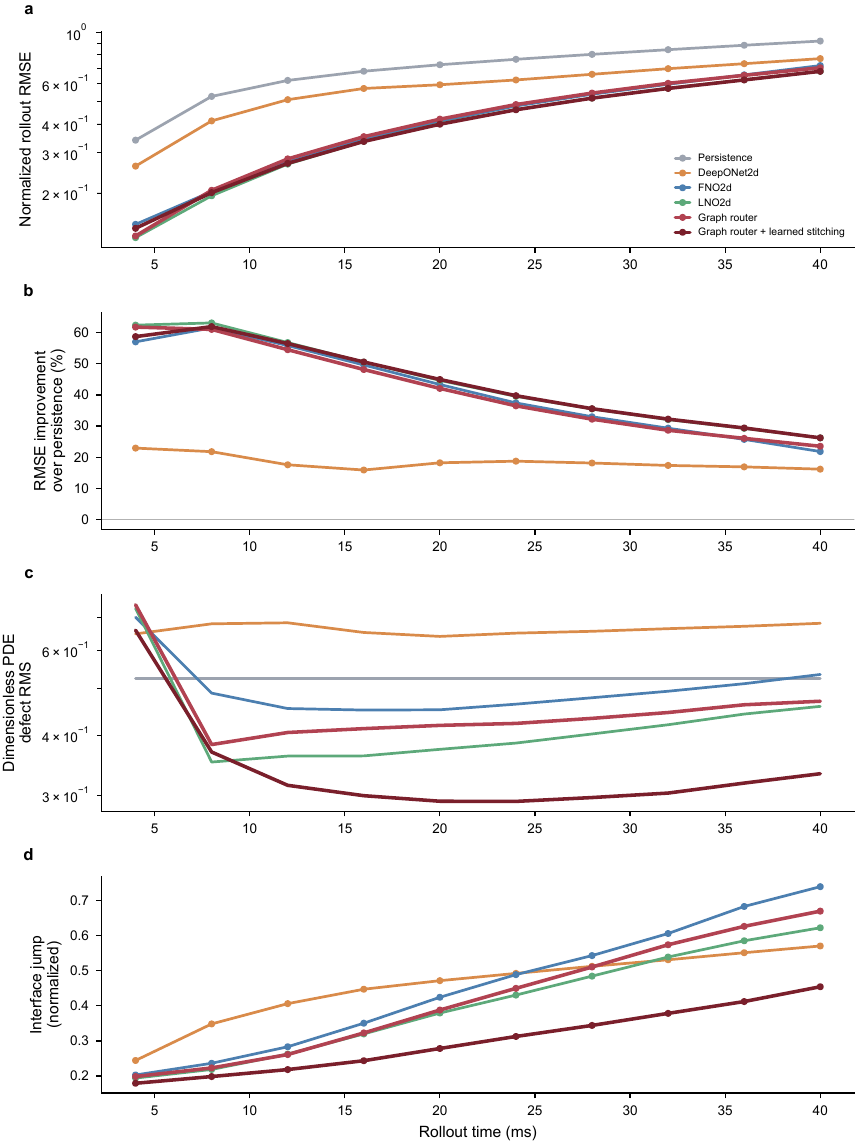}
  \caption{\textbf{Seam-aware dispatch comparison.} Rollout RMSE, improvement over persistence, PDE defect, and interface jump for the base and seam-aware policies.}
  \label{fig_5}
\end{figure}

\paragraph{Dispatch and autoregressive rollout}
Figure~\ref{fig_2} shows the specialist assignments produced by RCNO for three test cases. The spatial dispatch pattern changes substantially from case to case, indicating that routing is not a fixed partition of the domain. The comparison in Figure~\ref{fig_5} further illustrates the effect of using the orchestrated policy versus a single fixed operator.

\begin{figure}[h!]
  \centering
  \begin{subfigure}{0.49\linewidth}
    \includegraphics[width=\linewidth]{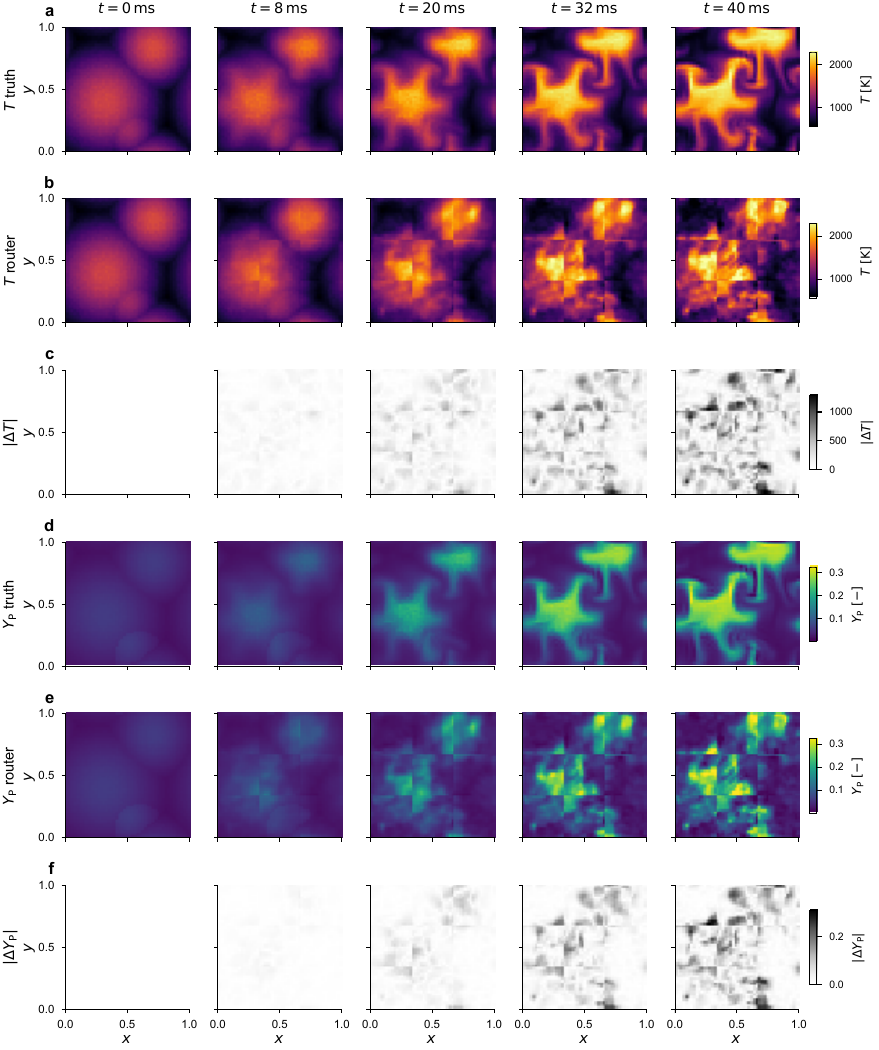}
  \end{subfigure}
  \hfill
  \begin{subfigure}{0.49\linewidth}
    \includegraphics[width=\linewidth]{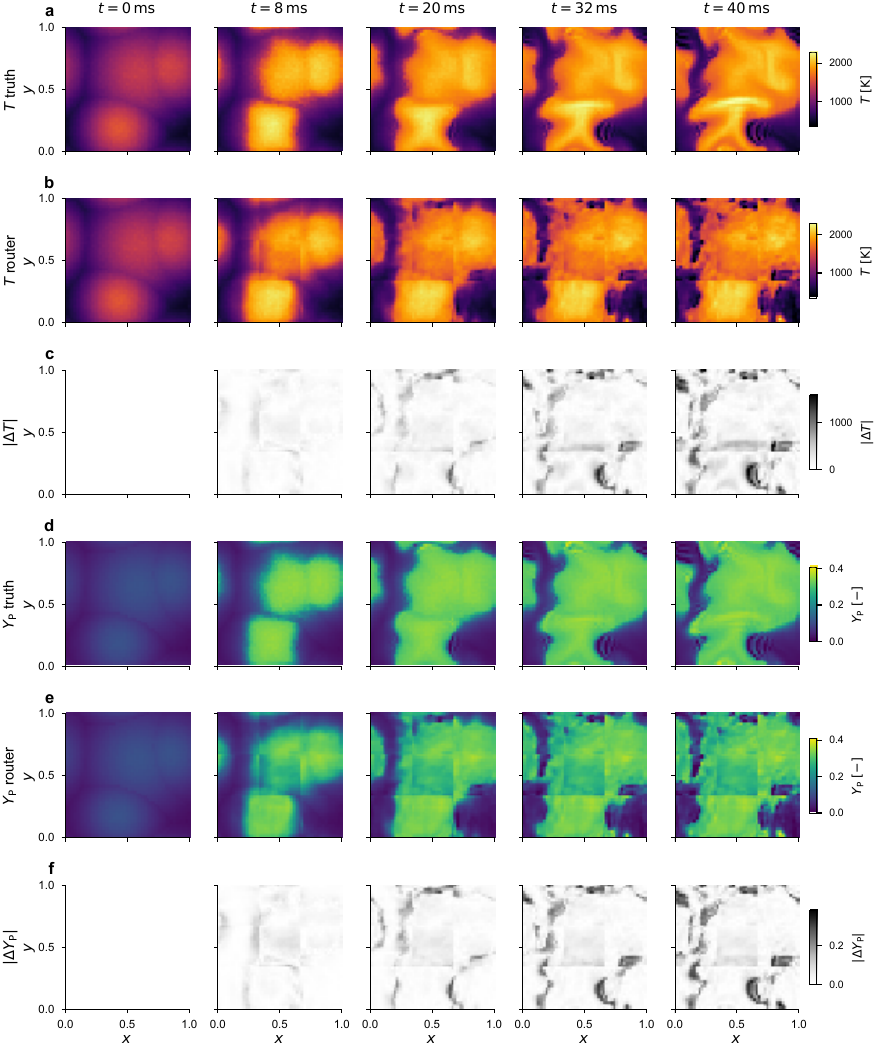}
  \end{subfigure}
  \caption{\textbf{Reacting-flow routed rollout.} Ground truth, orchestrated prediction, and absolute error for representative slow and fast cases.}
  \label{fig_3}
\end{figure}

In the baseline $3\times3$ decomposition, learned patch models substantially outperform persistence. The graph orchestrator tracks the strongest fixed policies in rollout error and gives a favourable PDE-defect profile (Figure~\ref{fig_5}). The routed rollouts in Figure~\ref{fig_3} show that the error concentrates near moving fronts and grows substantially in the faster case.
 
\paragraph{Decomposition granularity}
 
\begin{figure}[h!]
  \centering
  \includegraphics[width=0.40\linewidth]{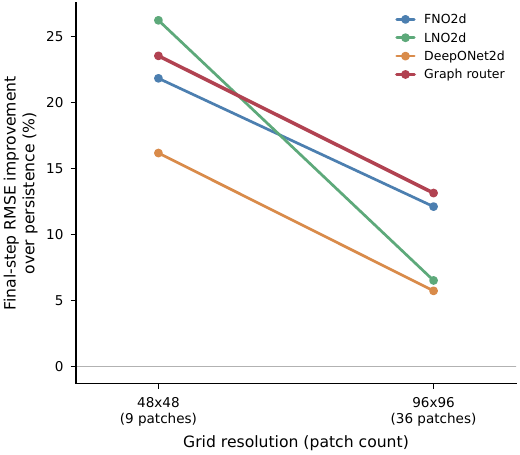}
  \caption{\textbf{Reacting-flow decomposition sensitivity.} Final-step RMSE improvement over persistence for fixed-architecture policies and the graph orchestrator at two patch counts.}
  \label{fig_4}
\end{figure}

The baseline reacting-flow configuration uses a $48\times48$ grid decomposed into 9 (i.e., $3\times3$) patches of size $16\times16$. We repeat the pipeline at $96\times96$ with a $6\times6$ decomposition while keeping the local patch size fixed (Figure~\ref{fig_4}). At the finer decomposition, the graph orchestrator becomes the most accurate, with a final improvement over persistence of $13.1\%$ versus $12.1\%$ for FNO, $6.5\%$ for LNO, and $5.7\%$ for DeepONet. The important observation is not simply the ordering of four models: the relative value of conditional composition increases as the domain contains more independently evolving local regions.
 
\paragraph{Seam-aware dispatch}
\label{sec:seam-ablation}
 
Because independently dispatched patches need not agree at their interfaces, we evaluate the seam-aware correction introduced in Eq.~\eqref{eq:coupled-update} and supervised through Eq.~\eqref{eq:interface-loss}. A correction network without interface supervision reduces the PDE defect but increases the final NRMSE and does not reduce the measured seam jump. When the interface penalty is explicitly weighted, mean jump falls by $28.5\%$, while final NRMSE and PDE defect improve by $3.5\%$ and $29.3\%$, respectively, relative to the base router. This shows that interface continuity is not a free consequence of accurate local prediction; it must be explicitly represented in the global composition objective.
 
\begin{table}[h!]
  \centering
  \caption{\textbf{Seam-aware reacting-flow ablation.} Final-horizon $10$-step rollout; mean $J$ is the interface jump associated with Eq.~\eqref{eq:interface-loss}, and lower values are better throughout.}
  \label{tab_2}
  \begin{tabular}{lrrr}
    \toprule
    Policy & Final NRMSE & Final PDE defect & Mean $J$\\
    \midrule
    Graph router & 0.7030 & 0.4700 & 0.4222\\
    + Stitcher, $\lambda_{\mathrm{interface}}=0$ & 0.7358 & 0.3741 & 0.4311\\
    + Stitcher, $\lambda_{\mathrm{interface}}=0.1$ & \textbf{0.6782} & \textbf{0.3325} & \textbf{0.3021}\\
    \bottomrule
  \end{tabular}
\end{table}
 
\section{Experiment II: Cahn--Hilliard--Navier--Stokes flow}
\label{sec:chns}
 
\paragraph{System, dataset, and regime features}
 
\begin{figure}[h!]
  \centering
  \begin{subfigure}{0.49\linewidth}
    \includegraphics[width=\linewidth]{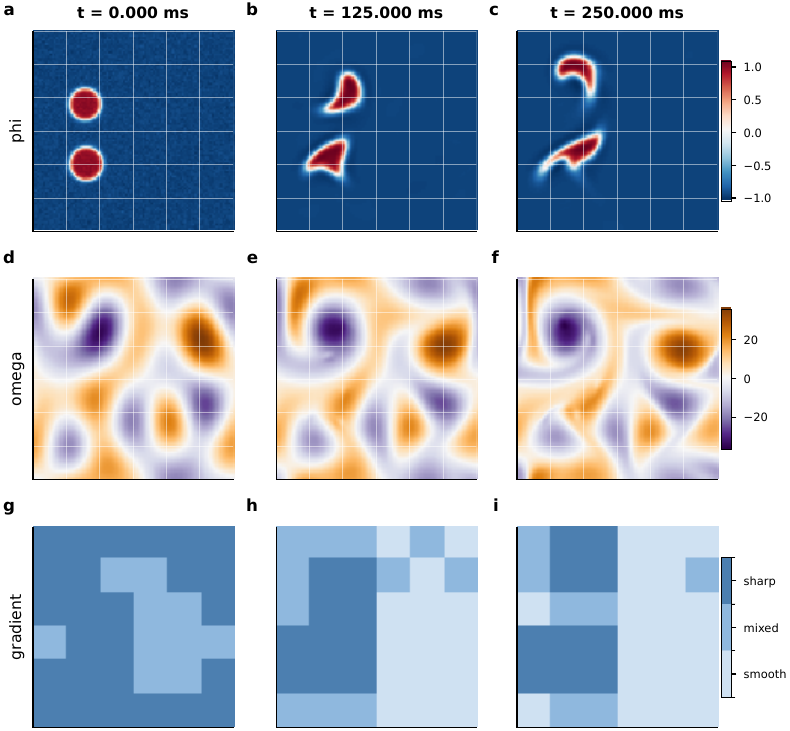}
  \end{subfigure}
  \hfill
  \begin{subfigure}{0.49\linewidth}
    \includegraphics[width=\linewidth]{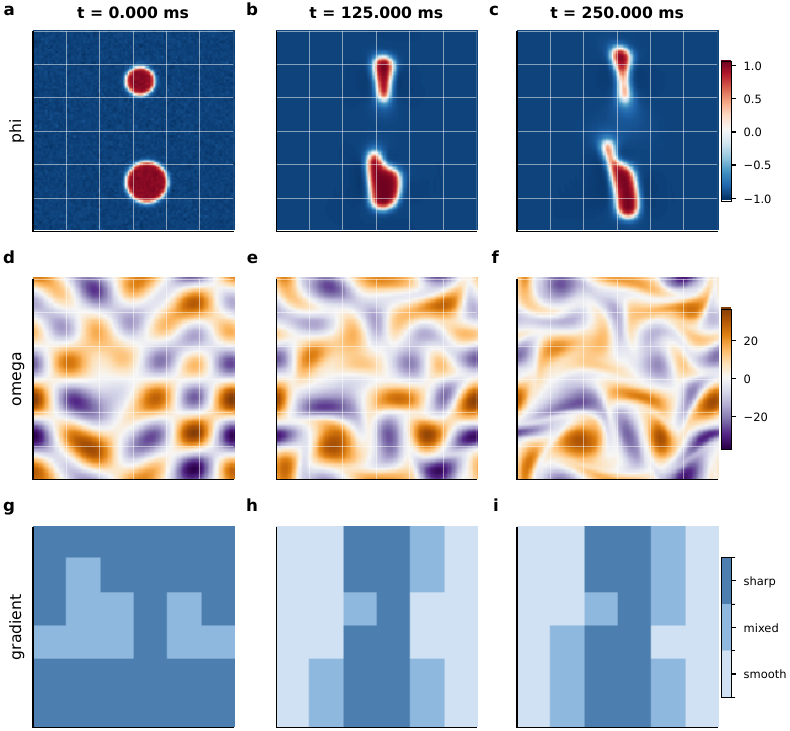}
  \end{subfigure}
  \caption{\textbf{CHNS benchmark.} Fast (left) and slow (right) dynamics. Rows $a$ and $b$ show phase field and vorticity, respectively; row $c$ shows the regime assigned to each patch.}
  \label{fig_7}
\end{figure}

The second benchmark is a multiphase, phase-field-based simulation, chosen to be drastically different from the reactive flow \cite{qiu2022physics,roccon2023phase,padhan2025cahn,zimmermann2020calculation}. The state of the system is represented by:
\begin{equation}
  \mathbf q_{\mathrm{CHNS}}=(\phi,\omega),
  \label{eq:chns-state}
\end{equation}
with velocity reconstructed from vorticity through a periodic streamfunction. The governing equations are
\begin{equation}
  \begin{aligned}
    \partial_t\phi+\mathbf u\cdot\nabla\phi
      &=M\nabla^2\mu,
    \qquad
    \mu=a(\phi^3-\phi)-\kappa\nabla^2\phi,\\
    \partial_t\omega+\mathbf u\cdot\nabla\omega
      &=\nu\nabla^2\omega+
        \sigma\nabla\times(-\phi\nabla\mu).
  \end{aligned}
  \label{eq:chns}
\end{equation}
Reference trajectories use a pseudo-spectral semi-implicit integrator, with the stiff Cahn--Hilliard and viscous terms treated implicitly and advection and capillary forcing treated explicitly. The headline dataset uses a $96\times96$ grid and a $6\times6$ decomposition into $16\times16$ patches. Cases independently vary mobility and interface width, producing combinations of slow/fast relaxation and diffuse/sharp interfaces (Figure~\ref{fig_7}).
   \begin{figure}[h]
    \centering
    \includegraphics[width=0.5\linewidth]{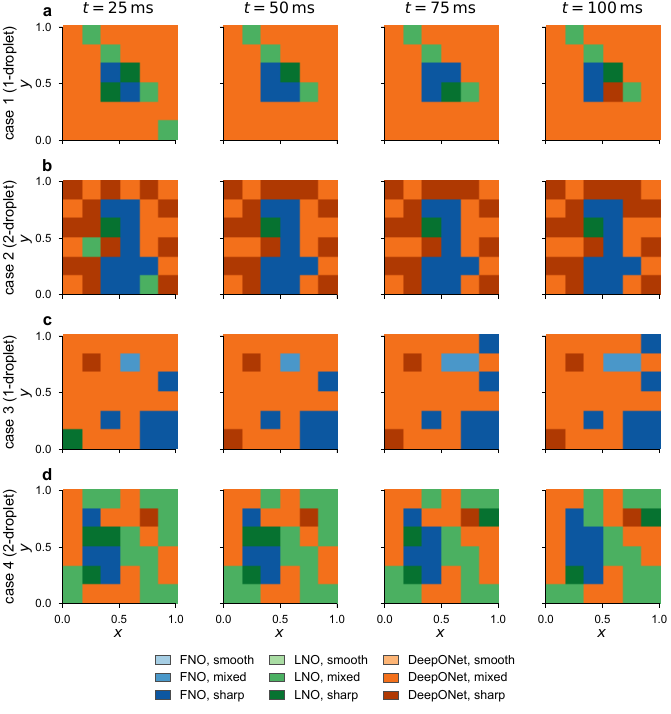}
    \caption{\textbf{CHNS expert dispatch.} Specialist assignments for four test cases on a $96\times96$ grid with $6\times6$ patches of $16\times16$ cells at $t=25,50,75,100$~ms.}
    \label{fig_6}
  \end{figure}
The implemented patch features are
\begin{equation}
  \mathbf r_{\mathrm{CHNS}}
  =
  \begin{bmatrix}
    r_{\mathrm{interface}}\\
    r_{\mu}\\
    r_{\mathrm{capillary}}\\
    r_{\mathrm{flow}}
  \end{bmatrix}
  =
  \log\!\left(
    1+
    \begin{bmatrix}
      \max|\nabla_{\mathrm{idx}}\phi|\\
      \max|\mu_{\mathrm{idx}}|\\
      \max|\nabla_{\mathrm{idx}}\times(-\phi\nabla_{\mathrm{idx}}\mu)|\\
      \max|\omega|
    \end{bmatrix}
  \right),
  \label{eq:chns-features}
\end{equation}
with maxima taken over the cells of the patch, and the activity is
\begin{equation}
  a_{\mathrm{CHNS}}
  =
  \tfrac12
  \left(
    r_{\mathrm{interface}}
    +
    r_{\mathrm{capillary}}
  \right).
  \label{eq:chns-activity}
\end{equation}
 
\paragraph{Specialisation, rollout, and dispatch}
 
The CHNS test exhibits a weaker but still useful architecture split than reacting flow. In this case we observed that DeepONet is generally strongest in the low-activity regime, while FNO is strongest in the intermediate and high-activity regimes. 
 
\begin{figure}[h!]
  \centering
  \includegraphics[width=0.72\linewidth]{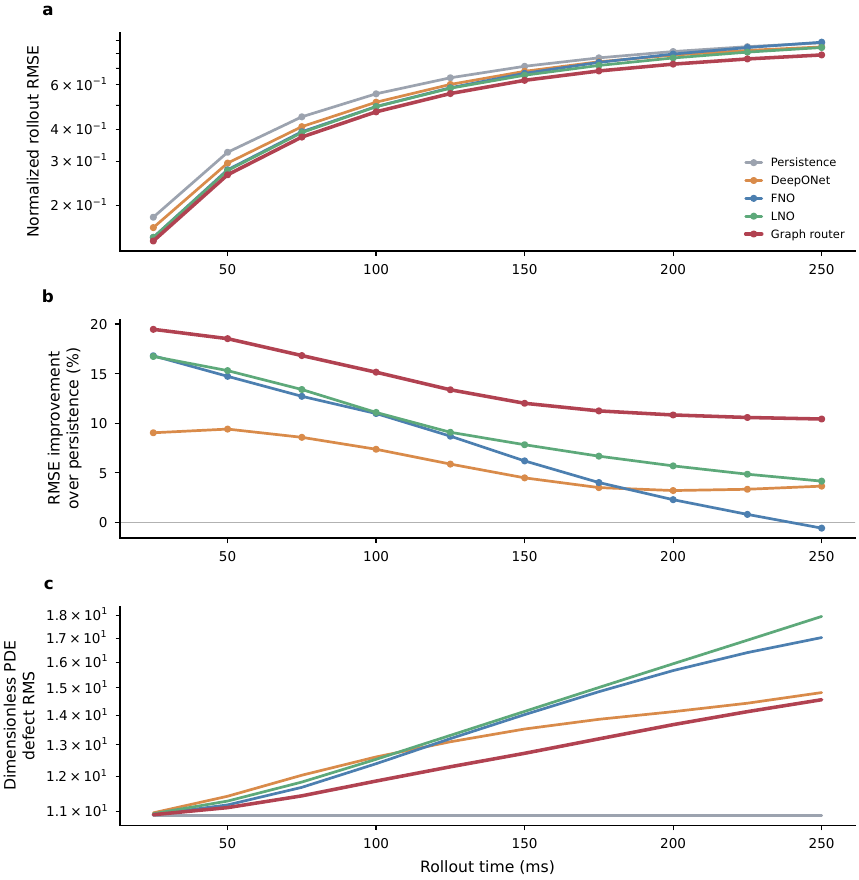}
  \caption{\textbf{CHNS autoregressive comparison.} Normalised RMSE, improvement over persistence, and dimensionless PDE defect across the rollout horizon.}
  \label{fig_8}
\end{figure}

The graph orchestrator has the lowest NRMSE at every plotted horizon (Figure~\ref{fig_8}) and reaches a final NRMSE of $0.7878$, compared with $0.8429$ for the strongest fixed-architecture case and $0.8794$ for persistence (Table~\ref{tab_3}). We also observe that, among learned surrogates, it gives the lowest final PDE defect. 
 
\begin{figure}[!h]
  \centering
  \includegraphics[width=0.5\textwidth]{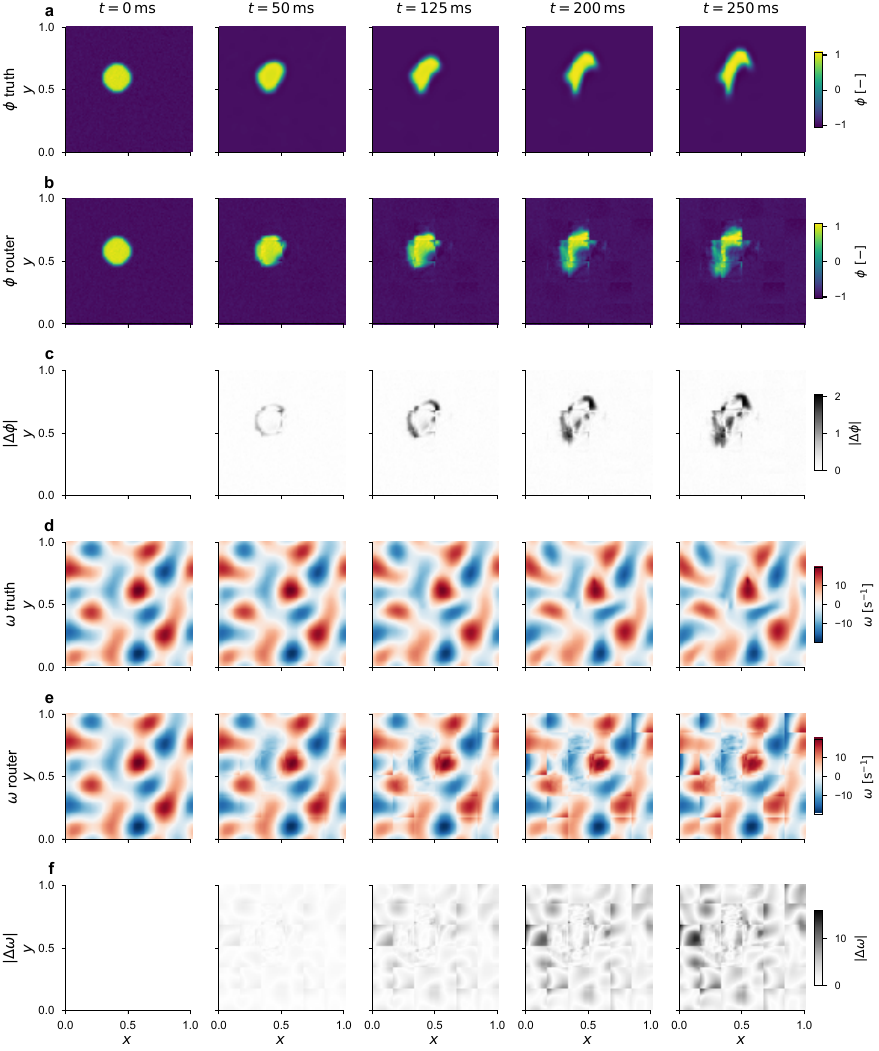}
  \caption{\textbf{CHNS routed rollout.} Phase field and vorticity ground truth, prediction, and absolute error at $t=0,50,125,200,250$~ms.}
  \label{fig_9}
\end{figure}

The sparse dispatch maps in Figure~\ref{fig_6} use seven of the nine expert nodes and span all three architectures. DeepONet specialists form much of the background, FNO specialists appear in compact structures associated with evolving interfaces, and LNO specialists occupy additional interface-adjacent regions. A routed rollout is shown in Figure~\ref{fig_9}. 

\begin{table}[h!]
  \centering
  \caption{\textbf{CHNS final-horizon results.} Performance at $96\times96$ with a $6\times6$ decomposition; lower values are better for NRMSE and PDE defect.}
  \label{tab_3}
  \begin{tabular}{lrrr}
    \toprule
    Policy & Final NRMSE & Final PDE defect & Improvement over persistence\\
    \midrule
    Persistence & 0.8794 & \textbf{10.8926} & $0.00\%$\\
     FNO & 0.8846 & 17.0196 & $-0.59\%$\\
     DeepONet & 0.8473 & 14.8236 & $3.65\%$\\
     LNO & 0.8429 & 17.9470 & $4.15\%$\\
    Graph orchestrator & \textbf{0.7878} & 14.5596 & $\mathbf{10.42\%}$\\
    \bottomrule
  \end{tabular}
\end{table}

\paragraph{Decomposition granularity}
\label{sec:chns-resolution}
 
\begin{figure}[!h]
  \centering
\includegraphics[width=0.50\linewidth]{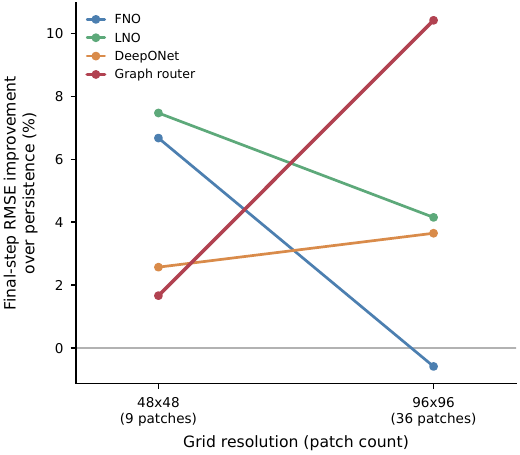}
  \caption{\textbf{CHNS decomposition sensitivity.} Final-step RMSE improvement over persistence for fixed-architecture policies and the graph orchestrator at two patch counts.}
  \label{fig_10}
\end{figure}

We repeated the CHNS test with a resolution of $48\times48$ and a $3\times3$ decomposition, and with a resolution of $96\times96$ and a $6\times6$ decomposition, while keeping the local patch size fixed (Figure~\ref{fig_10}).
 
  At the coarse decomposition, the graph router improves over persistence by $1.7\%$ and trails all direct applications of neural operator frameworks. At the finer decomposition the ordering changes: the router reaches a $10.4\%$ improvement, compared with $4.2\%$ for LNO, $3.6\%$ for DeepONet, and $-0.6\%$ for FNO. As in reacting flow, application of RCNO becomes more useful when the global domain contains more local regions over which different operator choices can be made. This trend is consistent with the greater routing flexibility available when more local regions are present.

\section{Conclusion}
\label{sec:conclusion}
 
We introduced the regime-aware coupled neural orchestrator (RCNO), a framework for dynamically composing independently trained local operator specialists over heterogeneous multiphysics domains. Instead of applying a single neural operator uniformly across the entire system, RCNO decomposes the state into local patches, characterises their current dynamical activity, and uses a graph orchestrator to decide which specialist should act in each region. The specialist bank combines the FNO, LNO, and DeepONet architectures across low-, intermediate-, and high-local-activity regimes. Expert representations first exchange query-conditioned messages, then are scored via edge-biased cross-attention. Sparse local predictions are assembled into a global state and projected onto system-specific physical constraints.
 
Across reacting flow and CHNS, the experiments show that useful local specialisation exists and that conditional composition becomes increasingly valuable as the spatial decomposition becomes finer. Expert-to-expert message passing improves both systems at finer levels of decomposition. However, it is not universally beneficial at coarse resolution, indicating that graph interactions should be interpreted as a conditional mechanism rather than an architectural default. The seam-aware study on reacting flow further shows that independently predicted patches introduce a second modelling problem at their interfaces. Explicit interface supervision reduces discontinuity, PDE defect, and rollout error simultaneously, demonstrating that global composition is incomplete even when the individual local models are accurate: the way their predictions are connected is itself part of the learned physical model.
 
The central contribution is therefore a framework for learning their time-varying spatial assembly. This provides a route toward modular surrogate models in which different operator inductive biases can be deployed where they are most useful. At the same time, the orchestrator learns how those local models should cooperate to recover the global multiphysics evolution.

\newpage
\bibliographystyle{unsrtnat}
\bibliography{bibliography}
 
\end{document}